\documentclass[twocolumn]{article}

\usepackage[
    top   = 0.8in,
    left  = 0.8in,
    right = 0.8in,
    bottom= 0.8in
    ]{geometry}
\usepackage{microtype} 
\usepackage[english]{babel} 

\usepackage{titling} 
\usepackage{titlesec} 
\renewcommand\thesection{\Roman{section}} 
\renewcommand\thesubsection{\Roman{section}.\arabic{subsection}} 
\titleformat{\section}[block]{\large\scshape}{\thesection.}{1em}{} 
\titleformat{\subsection}[block]{\scshape}{\thesubsection.}{1em}{} 

\usepackage{abstract} 

\usepackage{amsmath}
\usepackage{amssymb} 
\DeclareMathSizes{11}{9}{8}{6}  
\DeclareMathSizes{12}{10}{9}{8} 

\usepackage{booktabs} 
\usepackage{caption} 
\usepackage[numbib]{tocbibind} 
\usepackage{cite} 

\usepackage[hidelinks]{hyperref} 

\title{%
    \vspace{-2\baselineskip}
    \Huge\bfseries
    Determining Individual Origin Similarity (DInOS): \\
    \huge\bfseries
    Binary Classification of Authors Using Stylometric Features
    }
\author{%
    Kingsland, A.,
    Fortin, D.,
    Cary, E.,
    Smith, S.,
    Pazdernik, K.,
    and
    Perko, R.\\
    \small
    Pacific Northwest National Laboratory
}
\date{October 25, 2019}

\renewcommand{\maketitlehookd}{%
    \begin{abstract}
        \noindent
        Author similarity and detection is an integral first step in detecting state-led disinformation campaigns in an automated fashion.
        Current detection techniques require an analyst or subject matter expert to hand-curate accounts.
        Stylometric features have a rich history in identifying authorship of unknown documents, but little exploration has been done to compare authors to one another.
    We have adapted a select handful of stylometric features for use in author similarity metrics, and show their ${>}{0.96}$~F-1 performance on a curated author classification task, across both traditional machine learning and deep learning models.
        These features should contribute to the expanding field of author similarity research, and expedite the process of detecting and mitigating large-scale social media disinformation campaigns.
    \end{abstract}
}

\begin{document}

\maketitle

\section{Introduction}
In 2016, the Russia-based Internet Research Agency (IRA) conducted a multifaceted information warfare campaign at the direction of the Russian government~\cite{senate_russian_116, ica_assessing_2017}.
By posting a massive volume of falsified and deceptive information across a spectrum of social media and other online platforms, IRA operatives worked to undermine Americans’ faith in the 2016 U.S. electoral process and government institutions, as well as exacerbate political polarization by exploiting cultural and societal divisions~\cite{senate_russian_116, mazarr_hostile_2019, diresta_tactics_2018}.
Ongoing examination of these influence campaigns show evidence that not only has Russia continued to use this tactic to support their policy narratives and objectives, but that other state and non-state actors are emulating these tactics for their own use~\cite{senate_russian_116, mazarr_hostile_2019}.
Since the events of 2016, Twitter~\cite{twitter_ira_2018} and other social media platforms such as (but not limited to) Instagram, Facebook, and Tumblr~\cite{senate_russian_116, diresta_tactics_2018, tumblr_public_2018} have collected and published evidence of the IRA’s efforts to manipulate and deceive audiences both inside and outside of the United States.
Detection of troll-like user accounts is a first step in uncovering larger state-sponsored adversaries that are attempting to erode public trust in democratic institutions and processes.
Uncovering these users, though, is currently a time- and labor-intensive process that relies on analysts, subject matter experts, or content moderators evaluating accounts by hand in an effort to discover adversary narratives and map troll dissemination networks.
Many accounts, in order to evade detection, engage in the process of “persona-building”, or posting otherwise harmless materials to help establish their footprint as a legitimate account before publishing their deceptive or manipulative payload~\cite{senate_russian_116}. This can make it difficult to detect accounts of interest using solely content-based metrics.

There is a large body of work in authorship attribution through use of stylometrics~\cite{malyutov_authorship_2006, pavelec_compression_2009, stamatatos_survey_2009, layton_authorship_2010, narayanan_feasibility_2012, rocha_authorship_2016, overdorf_blogs_2016, neal_surveying_2018}, a field which attempts to match documents of unknown origins to their author.
Stylometrics applies features like
frequency of specific punctuation and n-grams,
number of characters and words,
part of speech usage,
vocabulary diversity,
and spelling and grammatical errors.
Stylometry, however, has only just begun to be applied to the field of author similarity detection~\cite{stolerman_classify_2013, afroz_doppelganger_2014, almishari_stylometric_2014},
though it has been proposed as early as 1987 with PCA~\cite{burrows_word_1987, burrows_not_1992}.

We believe that by applying stylometric features to text and incorporating them into machine learning pipelines, we will be able to accurately predict certain behaviors or qualities of relevance to national security.
This would increase the speed and precision by which analysts may identify problematic accounts, and allow institutions to counter these information warfare campaigns in a more timely and effective manner.
Stylometrics are also inherently devoid of metadata and, as such, they are platform-agnostic; that is, a fully-realized model should be able to detect similar users across platforms.

In this work, we focus on binary classification of Twitter data as a proof-of-concept for the use of stylometrics on author similarity tasks;
we explore a custom troll/not-troll dataset, which shows high performance across multiple machine learning models.
We also highlight current data limitations for author similarity tasks,
and consider which models and stylistic features are most appropriate going forward.


\section{Troll / not-Troll Dataset}

We developed a binary classification dataset: troll authors and non-troll authors. For the positive (troll) authors, we used the dataset released by Twitter~\cite{twitter_ira_2018} which contained tweets the platform attributed to the IRA's information warfare campaign directed against the 2016 presidential elections.
The IRA dataset contains tweets from known IRA trolls, including many tweets of ``innocent'' persona-building as well as more nefarious and divisive tweets (narrative payloads). Overall this class had over 53 million English tweets across 2,560 users.
The negative (not-troll) authors were taken from a garden hose Twitter drip over five days of July 2019.
This class had over 21 million English tweets across over 125,000 users.

Overall we found this dataset to be adequate to showcase the performance of our novel author similarity features and to compare multiple analysis methods. However, due to several assumptions made on the part of the authors, some limitations may exist in the dataset.
We assumed that troll-like behavior was not present or otherwise negligible in the garden hose data. We also assumed that weighing the models' loss functions for the different classes would account for the extreme imbalance in the number of authors for each label, and that the imbalance in the number of tweets would not affect performance. The final assumption is that the two sources are similar enough to merge, even given the multi-year gap between them.


\section{Features}

\subsection{Readability}%
\label{sec:readability}

There are many instances where an author's writing ability and command of the English language can be informative. Non-native English speakers may have a more limited vocabulary~\cite{mokhtar_achieving_2010}, or make common grammatical and spelling errors~\cite{futagi_computational_2008}.
Intuition would also suggest that a human author would write at about the same level, regardless of the topic, platform, or user account that they were currently using.

We propose using the following readability metrics as features, all intended to (roughly) represent reading grade level from 0--12, in our author similarity task:
\begin{itemize}
    \item The Dale-Chall readability formula~\cite{chall_readability_1995} (DCRF):
        \small
            \begin{equation}
                \begin{split}
                    \text{DCRF} &= \\
                      & 0.1579 \left(\frac{\text{difficult words}}{\text{words}} \times{} 100 \right) \\
                    + & 0.0497 \left(\frac{\text{words}}{\text{sentences}} \right) \\
                    + & \begin{cases}
                          3.6365 & \text{if}\; \frac{\text{difficult words}}{\text{words}} > 0.05, \\
                          0      & \text{else}
                      \end{cases}
                    ,
                \end{split}
            \end{equation}
            \normalsize
            where ``easy'' words are from a 3,000-word pre-determined list~\cite{chall_readability_1995}, and ``difficult'' words are all those which are not easy.
        \item The Automated readability index~\cite{smith_automated_1967} (ARI):
            \small
            \begin{equation}
                \begin{split}
                    \text{ARI} &= 4.71 \left(\frac{\text{characters}}{\text{words}} \right) \\
                               &+ 0.5 \left(\frac{\text{words}}{\text{sentences}}\right) - 21.43
                \end{split}
            \end{equation}
            \normalsize
        \item The SMOG index~\cite{mclaughlin_smog_1969}:
            \small
            \begin{equation}
                \begin{split}
                    \text{SMOG} &= 1.043 \sqrt{\text{polysyllables} \times\frac{30}{\text{sentences}}} \\
                                &+ 3.1291
                    ,
                \end{split}
            \end{equation}
            \normalsize
            where ``polysyllables'' refer to words which have 3 or more syllables.
        \item The Gunning fog index~\cite{gunning_technique_1968} (GFI):
            \small
            \begin{equation}
                \begin{split}
                    \text{GFI} = 0.4\Bigl[ & \left(\frac{\text{words}}{\text{sentences}}\right) \\
                                               &+ 100\left(\frac{\text{complex}}{\text{words}}\right)\Bigr]
                    ,
                \end{split}
            \end{equation}
            \normalsize
            where ``complex'' words are those which are both polysyllables, as defined by SMOG, and are difficult words, as defined by DCRF.\@
        \item The Coleman-Liau index~\cite{coleman_computer_1975} (CLI):
            \small
            \begin{equation}
                \begin{split}
                    \text{CLI} &= 0.0588\left(\frac{\text{letters}}{\text{words}}\right) \\
                               &- 0.2996\left(\frac{\text{sentences}}{\text{words}}\right) - 15.8
                \end{split}
            \end{equation}
            \normalsize
        \item The Flesch Kincaid grade level~\cite{kincaid_derivation_1975} (FKGL):
            \small
            \begin{equation}
                \begin{split}
                    \text{FKGL} &= 0.39\left(\frac{\text{words}}{\text{sentences}}\right) \\
                                &+ 11.8\left(\frac{\text{syllables}}{\text{words}}\right) - 15.59
                \end{split}
            \end{equation}
            \normalsize
        \item The Linsear write formula~\cite{klare_assessing_1974} (LW):
            \small
            \begin{equation}
                \text{LW} = \frac{\text{LW}_r}{2} -
                \begin{cases}
                    1 & \text{if}\; \text{LW}_r > 20, \\
                    0 & \text{else}
                \end{cases}
                ,
            \end{equation}
            \normalsize
            where the provisional result, $\text{LW}_r$, is given by:
            \small
            \begin{equation}
                \text{LW}_r = \frac{2\times\text{brachysyllables} + 3\times\text{polysyllables}}{\text{sentences}}
                ,
            \end{equation}
            \normalsize
            where ``brachysyllables'' are words with 1 or 2 syllables, and ``polysyllables'' are words with 3 or more syllables.
        \item The text standard ($\tilde{\mu}$), as defined by the mode of the above tests:
            \small
            \begin{equation}
                \begin{split}
                    \tilde\mu = \text{mode}\bigl(&\text{DCRF}, \text{ARI}, \text{SMOG}, \text{GFI}, \\
                                                   &\text{CLI}, \text{FKGL}, \text{LW}\bigr)
               \end{split}.
            \end{equation}
            \normalsize
        We use the mode to filter out extreme values and accurately represent the models' consensus.
    \end{itemize}

    \subsection{Lexical}
    A person's lexicon and use thereof should remain consistent, regardless of topic.
We identified several lexical features to use in our author similarity model:

    \begin{itemize}
        \item Statistics on a user's post length; the minimum, mean, median, and maximum number of characters were each used as individual features.
        \item The minimum, mean, median, and maximum number of unique words in a post.
        \item The author's overall minimum, mean, and maximum word length.
        \item 1,000 features from what we are calling the ``Burrows' Z'' metric, described in detail in Section~\ref{sec:technical-details}, which effectively captures an author's unique lexical fingerprint, as compared to other authors in the corpus.
    \end{itemize}

    \subsection{Syntactic}
    As with lexical features, a person's use of language should remain consistent, even across varied topics.
    We calculated syntactic features for each post, and aggregated them at the author level in an attempt to capture this effect. Author aggregations were the minimum, mean, median, and maximum across all of their post features. We calculated the post features as:

    \begin{itemize}
        \item Punctuation density (number of punctuation tokens divided by total number of tokens)
        \item Word density
        \item Digit density
        \item Special token (e.g.\ emoji) density
        \item Stop word density
        \item Meaningful word density
        \item Parts of speech densities; that is, for every present part of speech (e.g. ``adverb'') the number of its occurrences divided by the number of words
    \end{itemize}

    We have provided exact definitions for each of these categories as outlined in Section~\ref{sec:technical-details}.

    \subsection{Technical Details}%
    \label{sec:technical-details}

    For all models, all text was converted to lowercase, and tokens were separated from the text based on whitespace. Non-alphabetic characters were treated as separate tokens.
    Token classification was done with the Python 3.6 \texttt{string}~\cite{python_string_2016} library:
    punctuation tokens were considered to be any with a first character present in \texttt{punctuation},
    word tokens with first character in \texttt{ascii},
    and digit tokens with first character in \texttt{digits}.
    Special tokens were defined as any token present that was neither punctuation, word, nor digit.
    Stop words were any word token present within the English list provided by \texttt{nltk}'s~\cite{bird_natural_2009} \texttt{corpus.stopwords}. Meaningful words were any word token not within the stop word list.
    Part of speech tags were calculated using \texttt{nltk}'s~\cite{bird_natural_2009} \texttt{pos\_tag}, with only word-labeled tokens passed to the parser.
    Words' syllables were calculated with the \texttt{pyphen}~\cite{kozea_pyphen_2008} package.
    Sentences were counted approximately, by summing the number of ``.'', ``?'', and ``!'' special characters in each message. We forced a minimum count of one sentence.

    We additionally developed a set of 1,000 lexical features (given in Appendix~I) adapted from Burrows' Delta method~\cite{burrows_delta_2002}, which has been routinely used in author identification tasks. We call this feature set Burrows' Z,
    defined as the element-wise Z-score for author $a$:

    \begin{equation}
        \mathbf{Z}_a = (\mathbf{X}_a - \boldsymbol{\mu}) \odot \left[\frac 1{\sigma_0}, \hdots, \frac 1{\sigma_n}\right]
        ,
    \end{equation}

    \noindent
    where $\boldsymbol{\mu}$ is the element-wise mean of all $\mathbf{X}_a$, $\boldsymbol{\sigma}$ is the element-wise standard deviation of the same (a vector from $\sigma_0$ to $\sigma_n$), $\odot$ is the element-wise multiplicative operator, and each $\mathbf{X}_a$ is defined by

    \begin{equation}
        \mathbf{X}_a =
            \begin{bmatrix}
                P_a(w_0) \\
                P_a(w_1) \\
                \vdots \\
                P_a(w_{n-1}) \\
                P_a(w_n)
            \end{bmatrix}
        ,
    \end{equation}

    \noindent
    where $P_a(w_i)$ is the fraction of author $a$'s words that are $w_i$. Each $\mathbf{X}_a$ is calculated with the $n$ most-common words in the corpus; in our case we chose the top 1,000.

    Calculating the Burrows' Z-score in this manner has several important side-effects. Firstly, the corpus is normalized by \textit{author} vectors, $\mathbf{X}_a$, so that all authors are equally weighted within $\boldsymbol{\mu}$, regardless of the number of posts in the corpus.
    Secondly, each feature, $P_a(w_i)$, is scaled by the standard deviation, $\sigma_i$, to better differentiate usage at the tails of the distribution.
    Finally, scaling by the mean and standard deviation of each feature means that all of the features lie on the same scale, regardless of the frequency of their use within the corpus.

    All features (readability, lexical, and syntactic) were de-skewed using the Yeo-Johnson~\cite{yeo_new_2000} transform, as implemented by \texttt{sklearn}~\cite{pedregosa_sklearn_2011}, and then scaled to a mean of 0 with a standard deviation of 1.


\section{Experiments}

\subsection{Statistical Models}
We built and optimized three independent machine learning models: random forest, support vector machine (SVM), and extreme gradient boost (XGBoost) decision trees, as implemented by \texttt{sklearn}~\cite{pedregosa_sklearn_2011} and \texttt{xgboost}~\cite{chen_xgboost_2016}, respectively.
All models used balanced class weights, such that each class contributed equally to the loss.

We tuned the model hyper-parameters on the troll/not-troll dataset using Ray/Tune~0.7.3~\cite{liaw_tune_2018} with 1,000 trials for each model. Each trial had parameters randomly selected from pre-determined distributions given in Table~\ref{tab:hyper-parameters-ranges}.
The final parameters were taken from the best models, as defined by macro average F1-score on validation data, and given in Table~\ref{tab:hyper-parameters-best}.

\begin{table}[ht]
    \small
    \centering
    \begin{tabular}{lllc}
        \toprule
        \textbf{Model} & \textbf{Parameter} & \textbf{Range} \\
        \midrule
        Random   & Estimators         & $\mathbb{Z} \in [20, 1000]$ \\
        Forest   & Max.\ features     & \{log, sqrt\}               \\
                 & Max.\ depth        & $\mathbb{Z} \in [10, 110]$  \\
                 & Min.\ sample leaf  & $\mathbb{Z} \in [1, 5]$     \\
                 & Min.\ sample split & $\mathbb{Z} \in [2, 10]$    \\
        \midrule
        SVM      & Kernel             & \{linear, rbf\}             \\
                 & *Penalty param.    & $\mathbb{R} \in [1\text{e}{-1}, 1\text{e}{3}]$ \\
        \midrule
        XGBoost  & Estimators         & $\mathbb{Z} \in [20, 1000]$ \\
                 & Max.\ depth        & $\mathbb{Z} \in [1, 20]$    \\
                 & *Learning rate     & $\mathbb{R} \in [1\text{e}{-4}, 1\text{e}{-1}]$ \\
                 & *Gamma             & $\mathbb{R} \in [1\text{e}{-4}, 1\text{e}{0}]$  \\
        \midrule
        Deep     & Batch size         & $\mathbb{Z} \in [8, 64]$   \\
        Learning & Nodes per layer    & $\mathbb{Z} \in [16, 128]$ \\
                 & Hidden layers      & $\mathbb{Z} \in [0, 16]$   \\
                 & *Learning rate     & $\mathbb{R} \in [1\text{e}{-5}, 1\text{e}{2}]$ \\
        \bottomrule
    \end{tabular}
    \caption{%
        Parameter ranges used during hyper-parameter tuning of all models. All parameters were sampled from a random uniform distribution except where noted with an asterisk~(*), which indicates a log-uniform distribution. Integer parameters are represented by the letter $\mathbb{Z}$, and real-valued numbers by $\mathbb{R}$. Square brackets indicate that all numbers within the range were considered, including both endpoints. Curly braces indicate that the possible values were a set of equally-likely non-numerical choices.
    }%
    \label{tab:hyper-parameters-ranges}
\end{table}

We constructed a fourth model, an ensemble voting classifier, using \texttt{sklearn}~\cite{pedregosa_sklearn_2011}.
The model used each of the previous three models as sub-models, each implemented with its best hyper-parameters and weighed by its tuned performance, as defined by macro average F1-score, given in Table~\ref{tab:results-troll}.
The voting was conducted with soft voting, meaning that each model's probability of a given prediction were taken into account during the voting.

\subsection{Deep Learning Model}

We created a deep learning model with densely-connected layers using PyTorch~1.2.0~\cite{paszke_automatic_2017}.
We used the ReLU activation between each layer. Our final activation was a sigmoid function, and our loss function was binary cross-entropy, which we weighed such that each category (e.g. ``troll'' or ``not-troll'') contributed equally to the loss, regardless of any data imbalance. We optimized using the Adam~\cite{kingma_adam_2014} optimizer, using betas of 0.9 and 0.999, epsilon $1\text{e}{-8}$, and weight decay 0.

We refined the model hyper-parameters using Ray/Tune~0.7.3~\cite{liaw_tune_2018} to perform a meta-parameter search on 32~(troll) trials with parameter selection criteria given by Table~\ref{tab:hyper-parameters-ranges}.
Trials were stopped early if the test loss rose above 1.0 or at 10 epochs, whichever occurred sooner.
We chose the final parameters, given in Table~\ref{tab:hyper-parameters-best} from the trial epoch with the highest macro average F1-score.

\begin{table}[ht]
    \small
    \centering
    \begin{tabular}{lllc}
        \toprule
                       &                    & \textbf{Best}  \\
        \textbf{Model} & \textbf{Parameter} & \textbf{Value} \\
        \midrule
        Random   & Estimators         & 346  \\
        Forest   & Max.\ features     & sqrt \\
                 & Max.\ depth        & 10   \\
                 & Min.\ sample leaf  & 4    \\
                 & Min.\ sample split & 4    \\
        \midrule
        SVM      & Kernel             & rbf  \\
                 & Penalty param.     & $1.0\text{e}{0}$ \\
        \midrule
        XGBoost  & Estimators         & 705  \\
                 & Max.\ depth        & 4    \\
                 & Learning rate      & $9.0\text{e}{-2}$ \\
                 & Gamma              & $3.4\text{e}{-1}$ \\
        \midrule
        Deep     & Batch size         & 61   \\
        Learning & Nodes per layer    & 118  \\
                 & Hidden layers      & 6    \\
                 & Learning rate      & $5.6\text{e}{-4}$ \\
        \bottomrule
    \end{tabular}
    \caption{%
        The best meta-parameters for each model, as optimized on the troll/not-troll dataset. The best models were defined as those having the highest macro average F1-score.
    }%
    \label{tab:hyper-parameters-best}
\end{table}


\section{Results and Discussion}

\begin{table}[ht]
    \small
    \centering
    \begin{tabular}{llll}
        \toprule
                       & \textbf{Macro}     & \textbf{Macro}  & \textbf{Macro}             \\
        \textbf{Model} & \textbf{Precision} & \textbf{Recall} & \textbf{F1-score} \\
        \midrule
        Random Forest & 0.99 & 0.90 & 0.94 \\
        SVM           & 0.91 & 0.92 & 0.92 \\
        XGBoost       & \textbf{0.99} & \textbf{0.93} & \textbf{0.96} \\
        Voting        & 0.99 & 0.90 & 0.94 \\
        Deep Learning & 0.96 & 0.91 & 0.93 \\
        \bottomrule
    \end{tabular}
    \caption{%
        Performance results for all tuned models on the troll/not-troll dataset. Precision can be thought of as the true positive rate, and recall as the detection rate. Both are reported as the macro average, which is the mean of both troll and not-troll performance (equally weighted). The F1-score is the harmonic mean of precision and recall. XGBoost out-performs all other models on all measures.
    }%
    \label{tab:results-troll}
\end{table}

Performance results across all models are given in Table~\ref{tab:results-troll}.
Extreme gradient boost (XGBoost) out-performs all other models across all metrics,
and all of the statistic-based models out-performed the deep learning model.
This is likely due to the deep learning model overfitting the data very quickly-- almost all parameters' F1-score peaked at only one epoch. Overfitting would likely be slowed by much smaller learning rates, but it is unlikely that the final performance would out-perform that of XGBoost.
For the troll/not-troll dataset, we recommend using the XGBoost model, as it is the most performant.

Our performance is quite high, despite our implementing several readability features in Section~\ref{sec:readability} in a crude fashion.
For instance, the SMOG index was normalized on texts of 30 sentences, a length which we were unable to match given the short-message format of Twitter.
We also made two concessions in favor of scalability: 
while matching against the ``easy'' word list for Dale-Chall and Gunning fog, we eschewed the requirement that words be compared against their base form.
We forwent accurate sentence delineation, instead electing to count sentences based on quantity of punctuation.
As we are measuring author similarity and not grade-level reading, we are not so much concerned with the accuracy of a given readability score, so much as consistency and reproducibility, which our features provide.
Indeed, the high performance of all models on the troll/not-troll dataset indicate that our feature selection was appropriately predictive for this data.

It is possible that our models over-performed due to the inherent structure of our data set, having been joined from two disparate sources.
For instance, the word ``epstein'' was trending in the garden hose data (8,171 uses) but not in the IRA dataset (202 uses), which could be used to create very accurate labels based on the time period difference, but not necessarily on the users.
Most words, however, did not appear to be representative of a given time frame.
On a surface level, these words can appear to be quite mundane, but still \textit{highly} predictive of troll-like behavior (i.e.\ ``sports,'' ``local,'' or ``news'').
Many stopwords, which are by definition not correlated to any particular time event, were also indicative of trolls (i.e.\ ``after,'' ``against,'' or ``over'').
Other stopwords were anti-correlated with troll-like behavior (i.e.\ ``dont,'' ``been,'' or ``ve''),
which we hypothesize is due to language mastery, specifically with regard to contractions.
Overall, these seemingly-benign words show some of the underlying structure of the troll-like narratives, and in a manner detectable by our features.
Our approach provides a solid foundation for author similarity detection, even with a noisy data source.


\section{Conclusions}
High-volume author similarity and detection techniques are critical, not only in research applications, but with grave national security consequences.
Through the course of this study, we showed how the study of stylometrics, mostly used for author attribution, could be successfully adopted for author similarity and detection tasks.
Our classification models were highly performant on our stlye-based features, independently of model selection and implementation.
Our results could be somewhat inflated due to the incongruence of our data set: for future work, we heartily recommend more balanced data all taken within the same context and time period.

Author similarity models should help not only prevent future adversary-sponsored attempts at election interference, but also in detecting other coordinated attempts to sow discord, undermine civic trust in government institutions and processes, and propagate adversarial policy objectives and narratives.
Stylometric features, specifically, are inherently well-suited to cross-platform research and threat detection; they are not limited to one social media website's metadata format.

As more robust datasets for author similarity emerge, we expect to see stylometric features increase in prominence. More complex features, such as bigrams, grammatical errors, and usage anomalies, should be investigated. We also hope to explore multilingual models in order to remove the current requirement of English-only text when threats exist world-wide.
Finally, with more refined data, features, and models, we would hope to accomplish true author detection tasks. In particular, pinpointing a specific human with multiple user accounts, rather than this first elementary step of binary classification.


\small
\bibliography{bib}
\bibliographystyle{dinos}

\end{document}


\section{Top 1,000 words in troll/not-troll dataset}
All word tokens used in the calculation of Burrows' Z.
The number of uses (\textit{counts}) are the number of times that token (regardless of case) appeared in our troll corpus.
The troll percentage is the percentage of usages by positively-labeled trolls. Trolls accounted for 72.00\% of the word tokens in the corpus.
We also show the percentage points from random chance for each word token.
Stopword tokens are indicated in \textbf{bold}.

\bigskip

\tablefirsthead{%
    \toprule
    word token & number of uses & troll percentage & points from random (72.00) \\
    \midrule
}
\tablehead{%
    \toprule
    token & count & troll \% & points \\
    \midrule
}
\tabletail{\bottomrule}
\tablelasttail{\bottomrule}

\centering
\footnotesize
\begin{supertabular}{%
    p{0.35\linewidth}p{0.15\linewidth}p{0.15\linewidth}p{0.15\linewidth}}
    foke             & 28,633    & 100.0 & +27.99  \\
    fukushima2015    & 14,575    & 100.0 & +27.99  \\
    fukushimaagain   & 9,318     & 100.0 & +27.99  \\
    blicqer          & 8,795     & 100.0 & +27.99  \\
    giselleevns      & 7,463     & 100.0 & +27.99  \\
    danageezus       & 7,210     & 100.0 & +27.99  \\
    blacklivesmatter & 15,236    & 99.94 & +27.93  \\
    topnews          & 16,425    & 99.92 & +27.92  \\
\end{supertabular}
